\documentclass{vgtc}                          

\ifpdf
  \pdfoutput=1\relax                   
  \usepackage{graphicx}                
  \DeclareGraphicsExtensions{.pdf,.png,.jpg,.jpeg} 
\else
  \usepackage{graphicx}                
  \DeclareGraphicsExtensions{.eps}     
\fi%

\graphicspath{{figures/}{pictures/}{images/}{./}} 

\usepackage{microtype}                 
\PassOptionsToPackage{warn}{textcomp}  
\usepackage{textcomp}                  
\usepackage{mathptmx}                  
\usepackage{times}                     
\usepackage{cite}                      
\usepackage{tabu}                      
\usepackage{booktabs}                  
\usepackage{url}  

\usepackage[dvipsnames]{xcolor}
\definecolor{softblue}{HTML}{75a9be}
\definecolor{lightred}{HTML}{e53542}
\newcommand{\theName}{\textit{{Modie Viewer}}}

\onlineid{0}

\vgtccategory{Research}

\vgtcinsertpkg

\title{Modie Viewer: Protein Beasts and How to View Them}

\author{Huyen N. Nguyen\thanks{e-mail: huyen.nguyen@ttu.edu}\\ %
        \scriptsize Texas Tech University %
\and Caleb Trujillo\thanks{e-mail: calebtru@uw.edu}\\ %
     \scriptsize University of Washington Bothell %
\and Tommy Dang\thanks{e-mail: tommy.dang@ttu.edu}\\ %
     \parbox{1.4in}{\scriptsize \centering Texas Tech University}}

\abstract{
Understanding chemical modifications on proteins opens up further possibilities for research on rare diseases. This work proposes visualization approaches using two-dimensional (2D) and three-dimensional (3D) visual representations to analyze and gain insights into protein modifications. In this work, we present the application of \theName{} as an attempt to address the Bio+MedVis Challenge at IEEE VIS 2022.
} 

\CCScatlist{
  \CCScatTwelve{Human-centered computing}{Visu\-al\-iza\-tion}{Visu\-al\-iza\-tion techniques};
  \CCScatTwelve{Human-centered computing}{Visu\-al\-iza\-tion}{Visualization design and evaluation methods}{}
}

\begin{document}



\maketitle

\section{Motivation} 
As proteins are made and processed in a cellular environment, they fold to create three-dimensional structures due to intramolecular and intermolecular forces. Proteins are composed of polypeptides, which are composed of covalently bonded amino acids.  While the sequence of amino acids can provide information to identify the peptides, it is the molecular properties of the functional groups on the amino acids that provide the forces. A residue on a protein is one amino acid and the sequence of these residues makes up the primary structure. The primary structure of the amino acids folds, due to forces, to produce secondary structures such as alpha helices, beta sheets, folds, and bends. These secondary structures in turn combine to form tertiary structures, energetically stable peptide structures, which can also form quaternary structures when peptides combine as subunits of a larger functional protein. To better visualize the post-translational modifications that can occur, we display the number of changes possible at the residue level in three-dimensional tertiary structures based on solved X-ray crystallography data and computational models, along with two-dimensional visualizations to support in-depth investigation and analysis. The solved structures allow the viewers to see the broader properties and patterns of a molecule or complex.

\section{Observations on An Existing Visualization}

\begin{figure}[h]
 \centering 
 \includegraphics[width=\columnwidth]{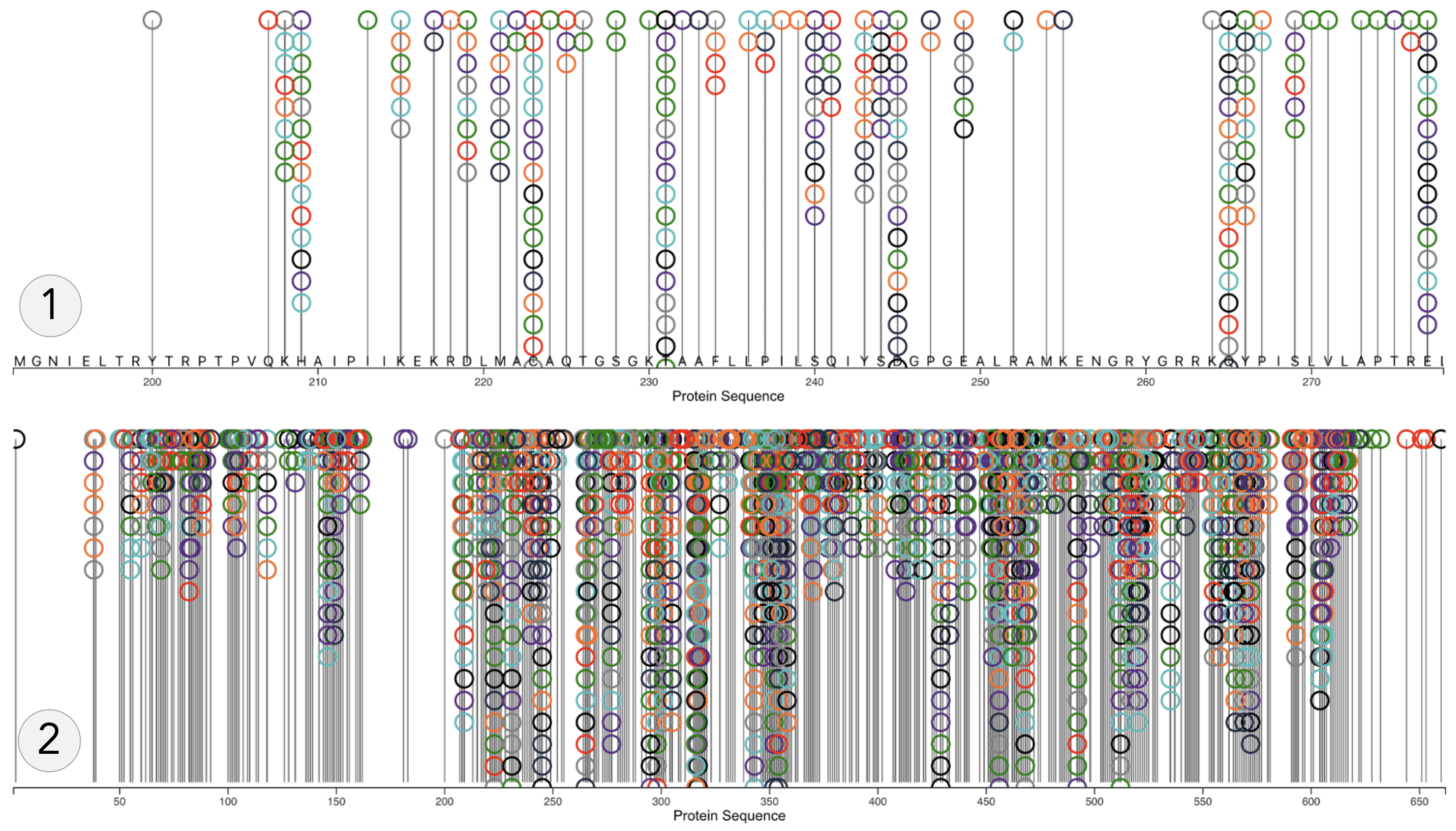}
 \caption{Existing visualization provided by Bio+MedVis Challenge at IEEE VIS~\cite{biovis_challenge}. The x-axis representing all residues in the sequence has been zoomed in (panel 1) and shows the full sequence (panel 2). The cluttered view in panel 2 is caused by overplotting.}
 \label{fig:redesign}
\end{figure}

The big question at the Bio+MedVis Challenge 2022~\cite{biovis_challenge} involves an existing visualization as presented in Figure~\ref{fig:redesign}: The vast number of modifications and modified residues cause a cluttered visualization as we go from an in-depth view (Figure~\ref{fig:redesign}-1) to a full view (Figure~\ref{fig:redesign}-2). 

The cluttered view in panel 2 is caused by a high information density with overplotted lines and circles. There are several directions for improvements for this visualization. First of all, better use of space should be considered: instead of using the one-dimensional vertical plotting upon a residue, horizontal space can be utilized as well. Second, there are various possibilities for color encoding that can be of use: representing modification type, classification of the type of post-translational modifications (PTMs), or mutation pathology. Finally, adding interactions could help in concentrating on the specific area for focus and context.

More clarification on visual encoding would be beneficial for this visualization. First is the usage of color for each circle representing a modification: whether the color stands for modification type, classification, or other information. Second is the underlying reasoning for the vertical order of modifications on each residue. Third is the meaning behind the height of the vertical line connecting all modifications upon one residue.

\section{Design Decisions on 2D Representations}
The visualization set in \theName{} includes (1) 3D Structures Viewer, (2) Classification on Protein Modifications in 2D, (3) Distribution of Modifications in 2D, and (4) Modification Types in 2D. This section specifically describes the design decisions for three 2D views; the following Section~\ref{sec:3D} will give an introduction to the 3D view.

\paragraph{\textbf{Basic Encodings}}
To support users with familiarity, the horizontal x-axis is kept to present all residues in the protein sequence. Each circle represents one modification.

\paragraph{\textbf{Utilizing Horizontal Space}}
The space above the x-axis is the one needed to be utilized more efficiently. In the (2) Classification view, each horizontal line represents one classification, indicating if the modification is a chemical derivative, an artefact, glycosylation, and so on. The circles are \textit{stacked horizontally} if there is more than one modification on a residue that belongs to the same classification.

\paragraph{\textbf{Color Encoding}}
A consistent color palette is used for all three visualizations. To enable both stacking and visibility, a semi-transparent opacity is applied for each circle representing modification. Mutations are marked with a black × mark.

\paragraph{\textbf{Ordering}}
We employ Hamming Distance~\cite{norouzi2012hamming} to measure the similarity between any two sequences of modifications. The ordering algorithm is applied for similarity sorting in the (2) Classification view and the (4) Modification type view. As the order on the x-axis is fixed, an efficient ordering on the vertical y-axis can greatly help users to discern any emerging patterns.

\paragraph{\textbf{User Interactions}}
The method of brushing and linking is an efficient approach to providing focus+context~\cite{nguyen2019eqsa}. For all 2D views, a bar chart showing classification distribution is appended under any main chart to give context, while the user can drag and resize the overlay sliding window to explore different parts of the sequence. In the (2) Classification view, the user can also drag and drop the classifications themselves for alternative orderings.

\section{3D Protein Structures}
\label{sec:3D}
For this challenge, we focus perceptual attention on the areas most likely to be modified in the context of the protein structure. We used visualization interactions to reduce the amount of visual processing and a simple color scheme to communicate both the variation among the residues and across the sample models. By using the 3Dmol~\cite{rego20153dmol} tools, users can rotate, zoom and tilt a molecule to view it from any angle. 

In the R environment, we were able to create a work flow by first, using accession numbers, downloading the PDB~\cite{pdb} files and AlphaFold files, selecting the highest resolution models for each X-ray crystallography structure, and then linking these files at the residue to modification data. Using r3Dmol, we rendered the structures and colorized the models as follows: White for modification is absent. Soft blue for 1-10 modifications possible. Light red for 11 or more modifications possible.

\section{\theName{} Platform}
\theName{} is available online at \url{https://huyen-nguyen.github.io/biomed-vis/}. 

\paragraph{Implementation} The 2D visualizations are developed with D3.js~\cite{bostock2011d3}. The 3D visualization is built with R using the packages of Protti~\cite{protti},  R3dmol~\cite{r3dmol}, Tidyverse~\cite{tidyverse}, Bio3d~\cite{grant2006bio3d} and 3Dmol~\cite{rego20153dmol}.

\section{Results and Discussion}
\begin{figure*}
 \centering 
 \includegraphics[width=\linewidth]{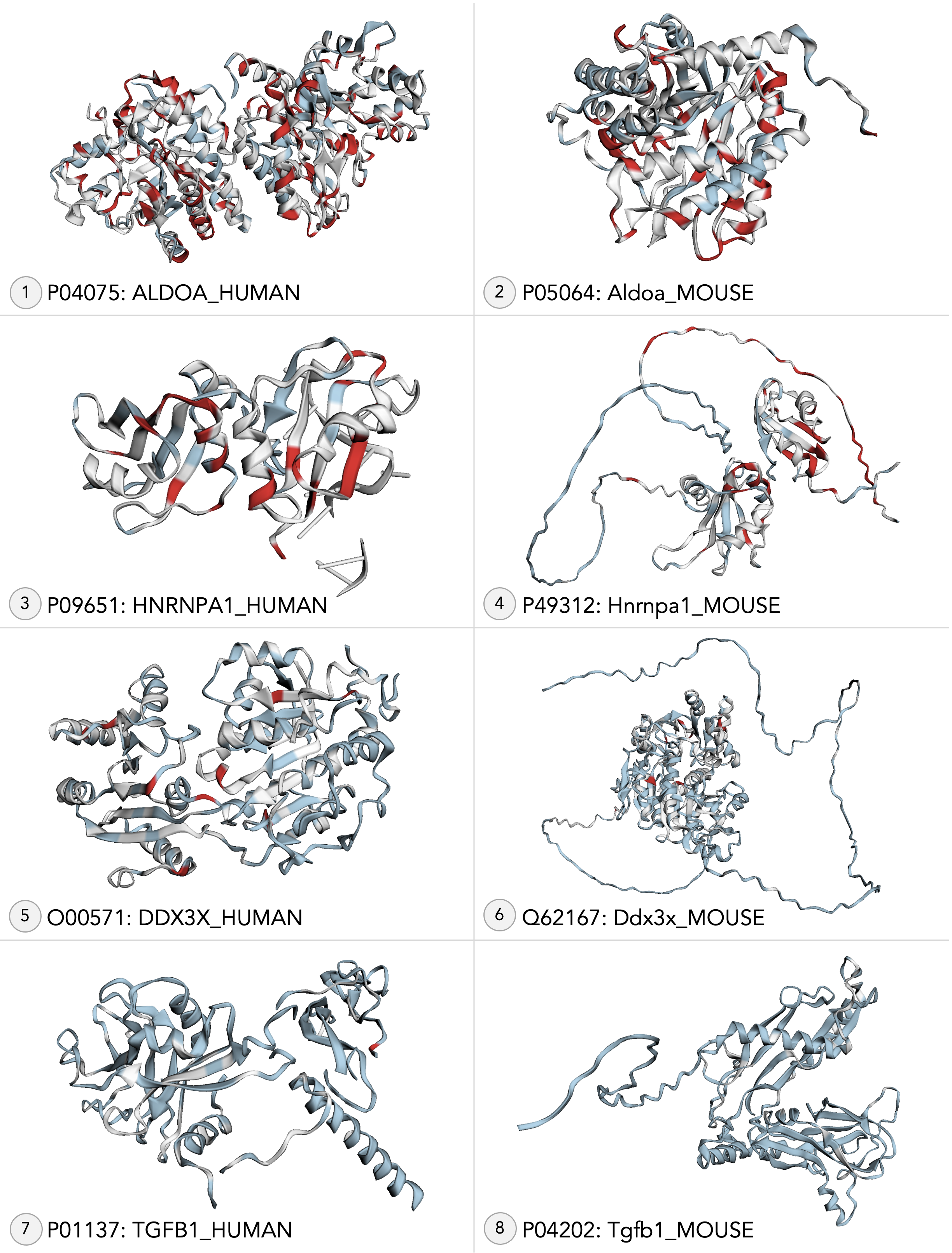}
 \caption{Eight 3D structures for four human proteins (ALDOA, DDX3X, HNRNPA1 and TGFB1) and their mouse orthologs. Each molecule is listed with its accession number and colorized for hot spots: White for modification is absent. \textcolor{softblue}{Soft blue} for 1-10 modifications possible. \textcolor{lightred}{Light red} for 11 or more modifications possible.}
 \label{fig:3D}
\end{figure*}

Eight 3D structures for four human proteins (ALDOA, DDX3X, HNRNPA1 and TGFB1) and their mouse orthologs are demonstrated in Figure~\ref{fig:3D}. From the 3D model results, ALDOA and its mouse ortholog have the largest number of modification occurrences, as the red areas are shown more frequently, followed by HNRNPA1 and its ortholog, then the DDX3X group and TGFB1 group. 

\begin{figure*}
 \centering
 \includegraphics[width=\linewidth]{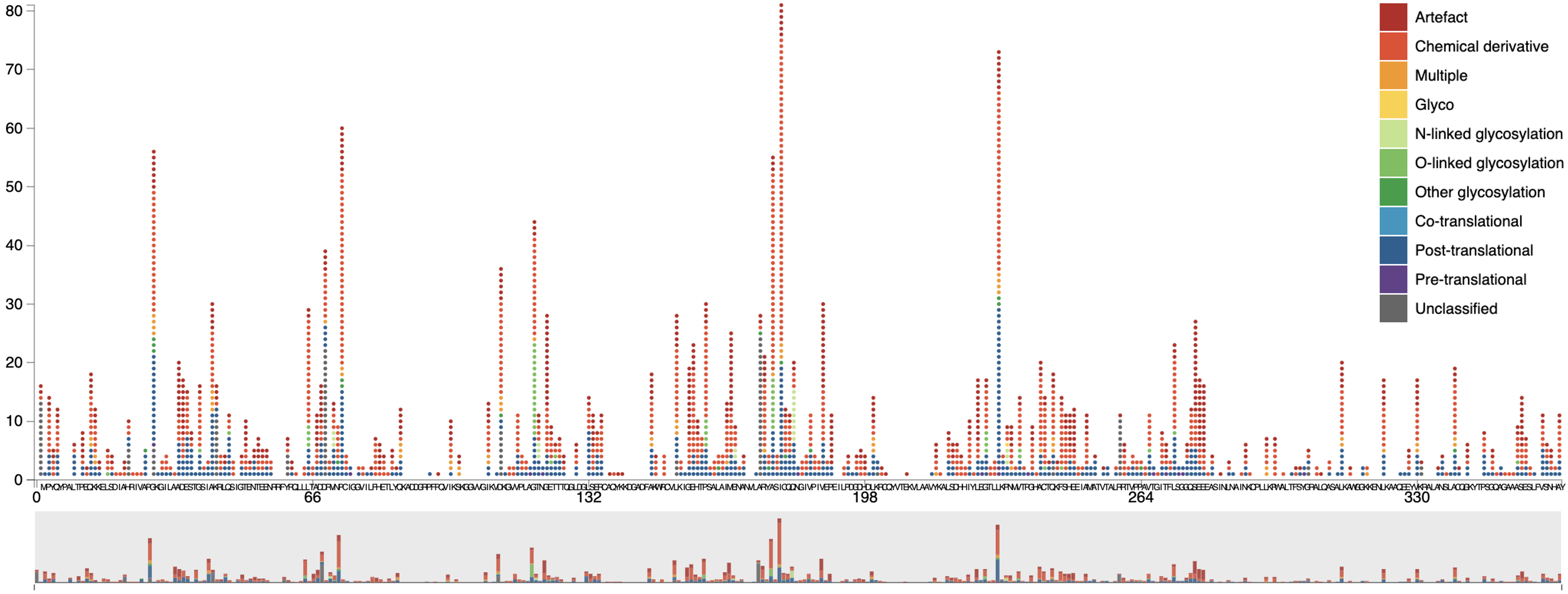}
 \caption{Modification Distribution on the protein sequence of ALDOA (Human), accession number P04075. At position 178, residue C (Cysteine, Cys) has the largest number of modifications on a site, at 81 modifications. Among 8 molecules within the dataset, this sample witnessed the highest number of modifications of 2,208 in total.}
 \label{fig:dist}
\end{figure*}

An overview of modification distribution is presented in Figure~\ref{fig:dist}, with the example of the protein sequence of ALDOA (Human), accession number P04075. This protein is known to be important in the rare disease of Hereditary nonspherocytic hemolytic anemia (HNSHA), the main feature of which is the premature destruction of red blood cells. At position 178, residue C (Cysteine, Cys) has the largest number of modifications on a site, at 81 modifications. Among 8 molecules within the dataset, this sample witnessed the highest number of modifications of 2,208 in total. In general, the most frequent residue to have modifications on is also C; the most common modification type is Oxidation, while that of classification is Chemical derivative.

\begin{figure*}
 \centering
 \includegraphics[width=\linewidth]{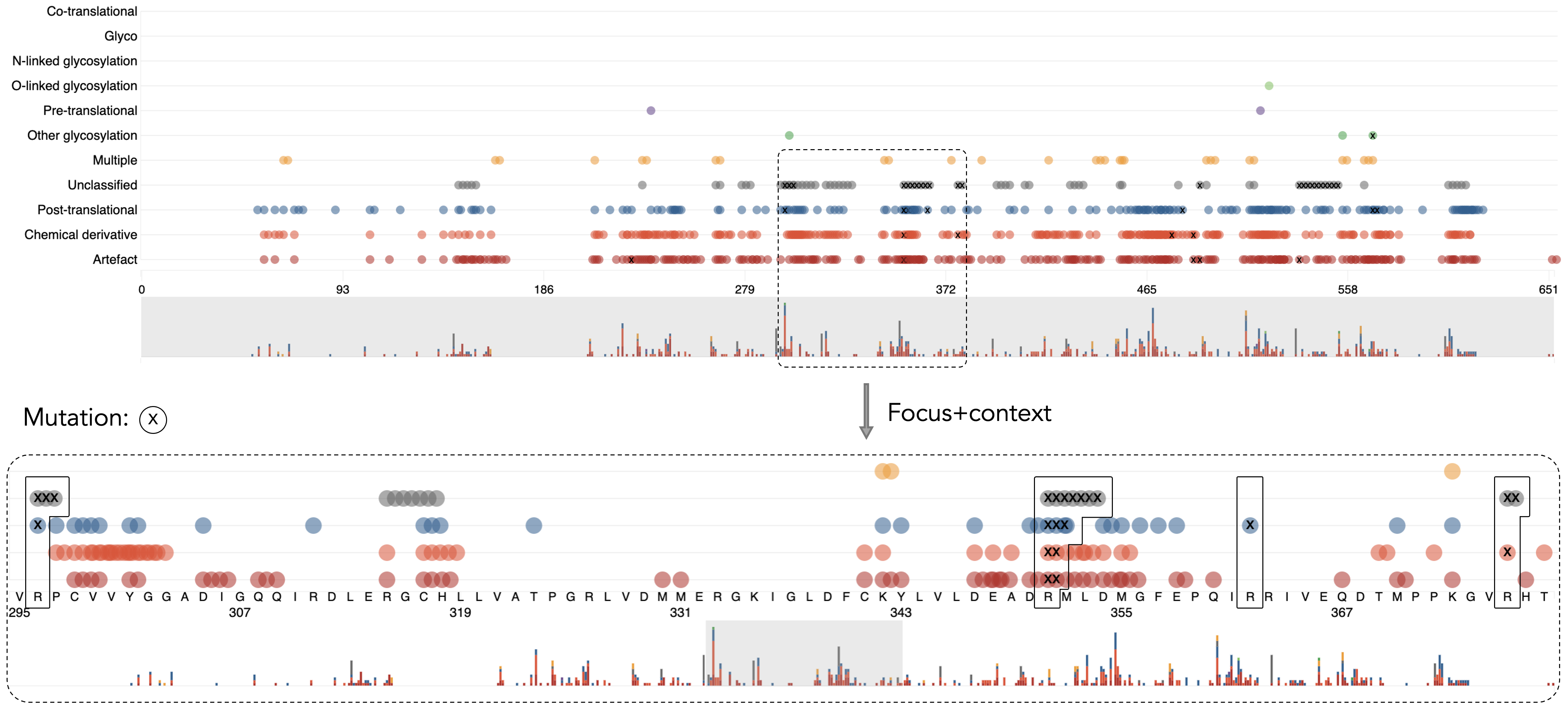}
 \caption{Classification on Protein Modifications. This example is based on the protein sequence of DDX3X (Human), accession number O00571. The majority classifications are Artefact, Chemical derivative, and Post-translational. In the zoomed-in part of the sequence, all mutations happen to occur on residue R (Arginine, Arg), an amino acid with electrically charged side chains – basic.}
 \label{fig:cls}
\end{figure*}

With the improvement in utilization of the horizontal space with stacking, the issue of overplotting is alleviated. Figure~\ref{fig:cls} demonstrates the classification of protein modifications with horizontal stacking. This visualization is based on the protein sequence of DDX3X (Human), accession number O00571. The majority classifications are Artefact, Chemical derivative, and Post-translational. With the similarity ordering, the classifications that have similar modification sequences are closer together.

Interactions help users examine further their view of interest. In the zoomed-in panel of the protein sequence using focus+context in Figure~\ref{fig:cls}, we can see that all mutations happen to occur on residue R (Arginine, Arg), an amino acid with electrically charged side chains – basic. The corresponding positions are 296, 351, 362, and 376. This observation could be invested in further research in terms of mutation characteristics and chemical modifications on proteins.

\begin{figure*}
 \centering
 \includegraphics[width=\linewidth]{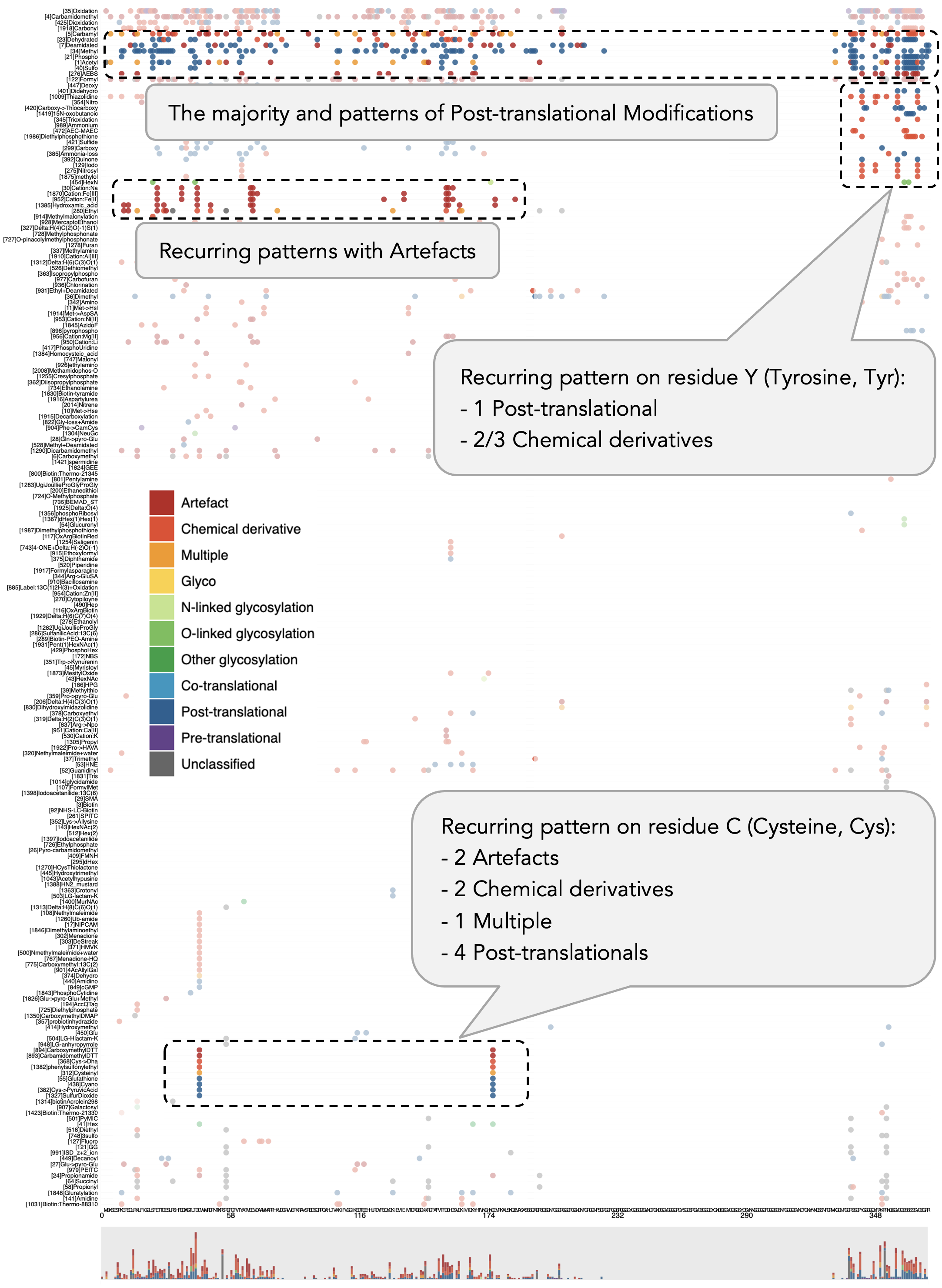}
 \caption{Modification Types on the protein sequence of HNRNPA1 (Human), accession number P09651. The recurring pattern appears on residue C at positions 43 and 175. This pattern contains two Artefacts, two Chemical derivatives, one Multiple, and four Post-translational modifications.}
 \label{fig:type1}
\end{figure*}

For a close-up view of modification types, Figure~\ref{fig:type1} presents modification types on the protein sequence of HNRNPA1 (Human), accession number P09651. This protein is considered important in relating to Amyotrophic lateral sclerosis, a rare disease on nervous system. This visualization consists of modification types on the y-axis and residues on the x-axis. There is a recurring pattern that appears on residue C at positions 43 and 175: This pattern contains two Artefacts, two Chemical derivatives, one Multiple, and four Post-translational modifications. Without a proper sorting on the vertical axis that groups similar modification types together, it is more challenging to discern a pattern like this one.

\section{Conclusion}
In this work, we propose the application of~\theName{}, a visualization platform utilizing two-dimensional (2D) and three-dimensional (3D) visual representations to understand where protein modifications are most likely to occur and gain insights from the distribution of such modifications in the protein sequence. The findings suggest directions for further research on mutation characteristics and a better visualization system for chemical modifications on proteins.

\bibliographystyle{abbrv-doi}

\bibliography{references}

\begin{thebibliography}{10}

\bibitem{biovis_challenge}
{Bio+MedVis Challenge \@ IEEE VIS}.
\newblock \url{http://biovis.net/2022/biovisChallenges_vis/}.
\newblock Accessed: 2022-09-09.

\bibitem{pdb}
Protein data bank.
\newblock \url{https://www.rcsb.org/}.
\newblock Accessed: 2022-09-09.

\bibitem{bostock2011d3}
M.~Bostock, V.~Ogievetsky, and J.~Heer.
\newblock D$^3$ data-driven documents.
\newblock {\em IEEE transactions on visualization and computer graphics},
  17(12):2301--2309, 2011.

\bibitem{grant2006bio3d}
B.~J. Grant, A.~P. Rodrigues, K.~M. ElSawy, J.~A. McCammon, and L.~S. Caves.
\newblock Bio3d: an r package for the comparative analysis of protein
  structures.
\newblock {\em Bioinformatics}, 22(21):2695--2696, 2006.

\bibitem{nguyen2019eqsa}
H.~N. Nguyen and T.~Dang.
\newblock {EQSA: Earthquake situational analytics from social media}.
\newblock In {\em 2019 IEEE Conference on Visual Analytics Science and
  Technology (VAST)}, pp. 142--143. IEEE, 2019.

\bibitem{norouzi2012hamming}
M.~Norouzi, D.~J. Fleet, and R.~R. Salakhutdinov.
\newblock Hamming distance metric learning.
\newblock {\em Advances in neural information processing systems}, 25, 2012.

\bibitem{protti}
J.-P. Quast, D.~Schuster, and P.~Picotti.
\newblock protti: an r package for comprehensive data analysis of peptide- and
  protein-centric bottom-up proteomics data.
\newblock {\em Bioinformatics Advances}, 2(1), 2022.

\bibitem{rego20153dmol}
N.~Rego and D.~Koes.
\newblock {3Dmol.js: molecular visualization with WebGL}.
\newblock {\em Bioinformatics}, 31(8):1322--1324, 12 2014. doi: {{%
10\hspace{.1pt}\discretionary{.}{%
}{.}\hspace{.4pt}1093\discretionary{/}{%
}{/}bioinformatics\discretionary{/}{%
}{/}btu829}}


\bibitem{r3dmol}
W.~Su and B.~Johnston.
\newblock r3dmol.
\newblock \url{https://swsoyee.github.io/r3dmol/}.
\newblock Accessed: 2022-09-09.

\bibitem{tidyverse}
H.~Wickham, M.~Averick, J.~Bryan, W.~Chang, L.~D. McGowan, R.~François,
  G.~Grolemund, A.~Hayes, L.~Henry, J.~Hester, M.~Kuhn, T.~L. Pedersen,
  E.~Miller, S.~M. Bache, K.~Müller, J.~Ooms, D.~Robinson, D.~P. Seidel,
  V.~Spinu, K.~Takahashi, D.~Vaughan, C.~Wilke, K.~Woo, and H.~Yutani.
\newblock Welcome to the {tidyverse}.
\newblock {\em Journal of Open Source Software}, 4(43):1686, 2019. doi: {{%
10\hspace{.1pt}\discretionary{.}{%
}{.}\hspace{.4pt}21105\discretionary{/}{%
}{/}joss\hspace{.1pt}\discretionary{.}{%
}{.}\hspace{.4pt}01686}}


\end{thebibliography}
\end{document}